\documentstyle[amssymb,aps,epsf,floats,twocolumn]{revtex}

\begin{document}

\def\jap{J.\ Appl.\ Phys.\ }
\def\cH{{\cal H}}
\def\m{{\cal M}}

\draft

\twocolumn[\hsize\textwidth\columnwidth\hsize\csname
@twocolumnfalse\endcsname

\title{Hole concentration in a diluted ferromagnetic semiconductor}
\author{Raimundo R.\ dos Santos$^{\, (1)}$, Luiz E.\ Oliveira$^{\, 
(2)}\!$, 
and J.\ d'Albuquerque e Castro$^{\, (1)}$}
\address{$^{(1)}$Instituto de F\' \i sica, Universidade Federal do Rio de Janeiro,
                 C.P.\ 68.528, 21945-970 Rio de Janeiro RJ, Brazil\\
$^{(2)}$Instituto de F\'{\i}sica, Unicamp, C.P. 6165,
13083-970 Campinas SP, Brazil}
\date{\today}
\maketitle
\begin{abstract}
We consider a mean-field approach to the hole-mediated ferromagnetism in
III-V Mn-based semiconductor compounds to discuss the dependence of the
hole density on that of Mn sites in Ga$_{1-x}$Mn$_x$As. The hole
concentration, $p$, as a function of the fraction of Mn sites, $x$, is
parametrized in terms of the product $m^*J_{pd}^2$ (where $m^*$ is the
hole effective mass and $J_{pd}$ is the Kondo-like hole/local-moment
coupling), and the critical temperature $T_c$. By using experimental data
for these quantities, we have established the dependence of the hole
concentration with $x$, which can be associated with the occurrence of a
reentrant metal-insulator transition taking place in the hole gas. We also
calculate the dependence of the Mn magnetization with $x$, for different
temperatures ($T$), and found that as $T$ increases, the width of the
composition-dependent magnetization decreases drammatically, and that the
magnetization maxima also decrease, indicating the need for
quality-control of Mn-doping composition in diluted magnetic semiconductor
devices.
\end{abstract}

]

Over the last few decades, a considerable amount of work has been devoted
to the understanding of electronic, optical and transport properties
of diluted magnetic semiconductors (DMS). Interest in these materials was
boosted in the early 1990's with the discovery of ferromagnetism in III-V
materials alloyed with transition elements like Mn\cite{Ohno92,Ohno96}.
Ferromagnetic semiconductors bring about the possibility of controlling
both spin and charge degrees of freedom, which, when combined with the
capability of growing low-dimensional structures, opens up exciting
new prospects for the production of spintronic devices. 
Potential applications include non-volatile memory
systems\cite{Ohno98a,Flederling99,Ohno99a,Ohno99c,Ohno00a} 
and quantum computing\cite{DiV99}. 

Especial attention has been focused on Ga$_{1-x}$Mn$_{x}$As alloys, which
exhibit very interesting magnetic and transport properties. Mn atoms have
five electrons in the 3d levels and two electrons in the 4s levels, and
their incorporation into a GaAs matrix plays two roles: they act both as
$S=5/2$ local moments, and as acceptors generating hole states in the
material. The equilibrium solubility of Mn atoms in GaAs is quite low,
being only of the order of $10^{19}$ cm$^{-3}$\cite{Hayashi97}. However,
with the use of molecular-beam epitaxy techniques at low temperatures,
several groups have recently succeeded in producing homogeneous samples of
Ga$_{1-x}$Mn$_{x}$As with $x$ as high as 0.071. It has been observed that
for $0.015\leq x \leq 0.071$ the systems become ferromagnetic, with
doping-dependent critical temperatures $T_{c}(x)$ reaching a maximum of
110 K for $x=0.053$\cite{Ohno01a}.

The appearance of a ferromagnetic state in these materials has been
attributed to an exchange coupling between the localized Mn moments
mediated by the holes, whose strength should depend on the hole
concentration $p$. In principle, one would expect that each Mn would
provide one hole, leading to a density of holes equal to that of the
magnetic ions. However, while an accurate determination of the hole
concentration is hindered by the anomalous Hall term, experimental data
indicate that $p$ is only a 15 to 30\% fraction of that of magnetic
ions\cite{Ohno01a,Ohno98b,Ohno99b,Dietl01}. The mechanism responsible for
the discrepancy between hole and Mn densities is not clear. As pointed out
by Matsukura {\it et al.}\cite{Ohno98b}, such discrepancy might be due to
compensation of Mn acceptors by deep donors such as As antisites, which
are known to be present at high concentration in low-temperature grown
GaAs\cite{Look91}. Another possibility would be the formation of
sixfold-coordinated centres with As (Mn$^{6{\rm As}}$), which would
compensate Mn atoms on substitutional Ga lattice sites\cite{vanEsch97}. As
a consequence, the relation between hole concentration and that of Mn has
not been so far theoretically established, which would be of great
interest for the design of new devices. Our main purpose here is to
present a quantitative analysis on this issue, based on a simple model for
the magnetic behaviour of these systems.

We adopt the generally accepted view that a given Mn ion interacts with
the holes via a local antiferromagnetic Kondo-like exchange coupling
$J_{pd}$ between their
moments\cite{Dietl01,Dietl97,Mac99a,Mac00a,Quinn00}. This interaction is
thought to lead to the polarization of the hole subsystem, which would
then give rise to an effective ferromagnetic coupling between the Mn
moments. Though there has been some debate as far as the details of the
above picture are concerned (e.g., whether or not such effective
interaction is well described by an RKKY term\cite{Mac00b,Litvinov01}),
there is an overall consensus on the fundamental role played by the
hole-mediated mechanism. At any rate, the approach we follow here does not
depend on the details of the effective Mn-Mn interaction.

We start with a Hamiltonian for the two coupled subsystems of the form 
\begin{equation}
{\cal H}={\cal H}_{{\rm Mn}}+{\cal H}_{{\rm h}}+J_{pd}\sum_{i,I}{\bf S}%
_{I}\cdot {\bf s}_{i}\ \delta \left( {\bf r}_{i}-{\bf R}_{I}\right) ,
\label{Ham1}
\end{equation}
where ${\cal H}_{{\rm Mn}}$ describes the {\em direct} ({\em i.e.,}
non--hole-mediated) antiferromagnetic exchange between Mn spins, ${\cal
H}_{ {\rm h}}$ describes the hole subsystem, and the last term corresponds
to the aforementioned Mn-hole exchange interaction, with ${\bf S}_{I}$ and
${\bf s} _{i}$ labeling the localized Mn spins ($S=5/2$) and the hole
spins ($s=1/2$ ), respectively. As a first approach, we neglect ${\cal
H}_{{\rm Mn}}$ and consider ${\cal H}_{{\rm h}}$ within a parabolic-band
effective-mass approximation; we comment below on more general
descriptions of ${\cal
H}_{{\rm h}}$.

Within a mean-field approximation, the Mn magnetization is given by 
\begin{equation}
M=N_{{\rm Mn}}g\mu _{{\rm B}}{\cal M}_{I}=n_{{\rm Mn}}g\mu _{{\rm B}}SB_{S}%
\left[ \left( {\frac{J_{pd}S}{2k_{{\rm B}}T}}\right) \ {\cal M}_{h}\right] ,
\label{M}
\end{equation}
where $n_{{\rm Mn}}=xn_{s}$ is the density of Mn ions, with $n_{s}$ being
the density of Ga lattice sites, ${\cal M}_{I}$ is the magnetization density
of the Mn ions, $g=2$ is the Mn Land\'{e} $g$-factor, and $B_{S}[\ldots ]$
is the Brillouin function. The magnetization density of the hole subsystem, $%
{\cal M}_{h}=\langle n_{\uparrow }-n_{\downarrow }\rangle $, is supposed to
be uniform within the length scale of magnetic interactions, so it can be \
calculated self-consistently by considering a Fermi sea of holes with
effective mass $m^{\ast }$, in the presence of the mean magnetic field
generated by the Mn ions; it is therefore given by 
\begin{equation}
{\cal M}_{h}=\lambda \ {\frac{m^{\ast }}{m_{e}}}J_{pd}\ x\ {\cal M}_{I}\ \
p^{1/3},  \label{mh}
\end{equation}
where $\lambda =6\left( 1/3\pi ^{2}\right) ^{2/3}\left( m_{e}/\hbar
^{2}a^{3}\right) $, $m_{e}$ is the free-electron mass; $a=5.65$ \AA\
is the GaAs lattice constant.

\begin{figure}[t]
\epsfxsize=3.5in
\centerline{\epsfbox{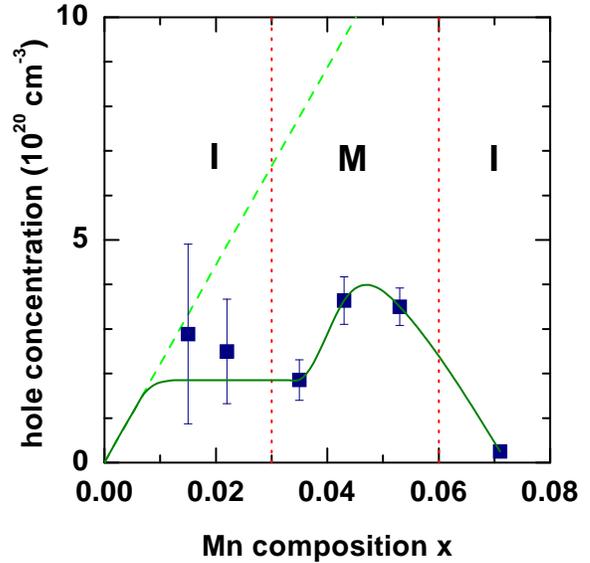}}

\caption{Theoretical results for the hole concentration as a function of
the fraction of Mn sites for a (Ga,Mn)As ferromagnetic alloy. The dashed
line corresponds to a hole concentration equal to that of Mn sites,
whereas the filled squares are the present mean-field result; the full
curve is obtained as reasoned in the text. I and M respectively denote
insulating and metallic phases.}

\label{px}
\end{figure}

The critical temperature as a function of the hole density and of the Mn
composition is obtained by linearizing the self-consistency relations
given by Eqs.\ (\ref{M}) and (\ref {mh}):
\begin{equation}
T_c={\frac{\lambda}{6k_{{\rm B}}}} S(S+1) \ 
\left[(m^*/m_e)J_{pd}^2\right]\ x\ p^{1/3}.
\label{tc}
\end{equation}
Specializing Eq.\ (\ref{tc}) to $S=5/2$, we write the hole concentration
as
\begin{equation}
p=\zeta \left\{ {\frac{T_{c}(x)}{\left[ (m^{\ast }/m_{e})J_{pd}^{2}\right] 
x}}\right\}^{3},  
\label{p}
\end{equation}
where $\zeta =5.29\times 10^{-16},$ in units such that $J_{pd}$ is given in
eV nm$^{3}$. 

In view of the uncertainty on the available experimental values for
$m^{\ast }$ and $J_{pd}$, and to the difficulties in obtaining accurate
estimates over a wide range of hole densities, $p$, the following strategy
is adopted. We first use the fact that Hall resistance measurements
\cite{Ohno99d} yield an unambiguous\cite{foot1} value of $p=3.5\times
10^{20} $ cm$^{-3}$ for the sample with $x=0.053$, for which $T_{c}=110$
K. We then take these values into Eq.\ (\ref{p}), to fit a value for the
product $(m^{\ast }/m_{e})J_{pd}^{2}=2.4\times 10^{-2}$ (eV
nm$^{3}$)$^{2}$. And, finally, we use this value, together with the
experimental transport data\cite{Ohno01a} for $T_{c}(x)$, to obtain $p$
over a wide range of $x$, shown as filled squares in Fig.\ \ref{px}. The
error bars in Fig.\ \ref{px} reflect the uncertainties in the
determination of $T_{c}(x)$, as displayed in Fig.\ 3(c) of Ref.\
\onlinecite{Ohno01a}. The adequacy of this procedure is illustrated in
Fig.\ \ref{px}: The calculated values for $p(x)$ lie below the
concentration of Mn ions, shown as a dashed line, in agreement with
experiment.  We also highlight in Fig.\ \ref{px} the boundaries of the
metal-insulator transitions (MIT's), as determined from resistivity
measurements\cite{Ohno01a}.  The present theoretical estimates for $p$ in
the insulating phases are based on the assumption that the localization
length in insulating samples, though finite, is significantly larger than
the length scale of magnetic interactions\cite{Dietl01}, in which case the
present mean-field approach is a good starting point.

Before accepting these estimates for $p(x)$ at face value, one should note
that a closer look at the experimental data for $T_c(x)$\cite{Ohno01a}
suggests a linear behaviour in the range of $x$ of the order 0.015-0.035
which would imply, through Eq.\ (\ref{p}), a constant $p$ in that range;
this constant behaviour, however should not prevail at low concentrations,
$x\to 0$, and presumably one should have $p\propto x\to 0$. These
considerations are incorporated in the full curve displayed in Fig.\
\ref{px}, which lies within the error bars of the calculated hole
concentrations.

Our theoretical estimates for $p(x)$ are therefore strongly suggestive of
$p(x)$ reaching a maximum value within the metallic phase. As a
consequence, all attempts to increase $T_{c}$ should be carried out for
samples in the metallic phase, for Mn concentrations about 0.05.  
Moreover, notwithstanding the considerable uncertainties\cite{foot2} in
the measurements of $p$, the data shown in Fig.\ \ref{px} are in
qualitative agreement with those obtained from Hall measurements by
Matsukura {\em et al.}\cite{Ohno98b}; as we discuss below, this is also
consistent with findings from recent photoemission spectroscopy
measurements\cite{photo}.

\begin{figure}[t]
\epsfxsize=3.5in
\epsfbox{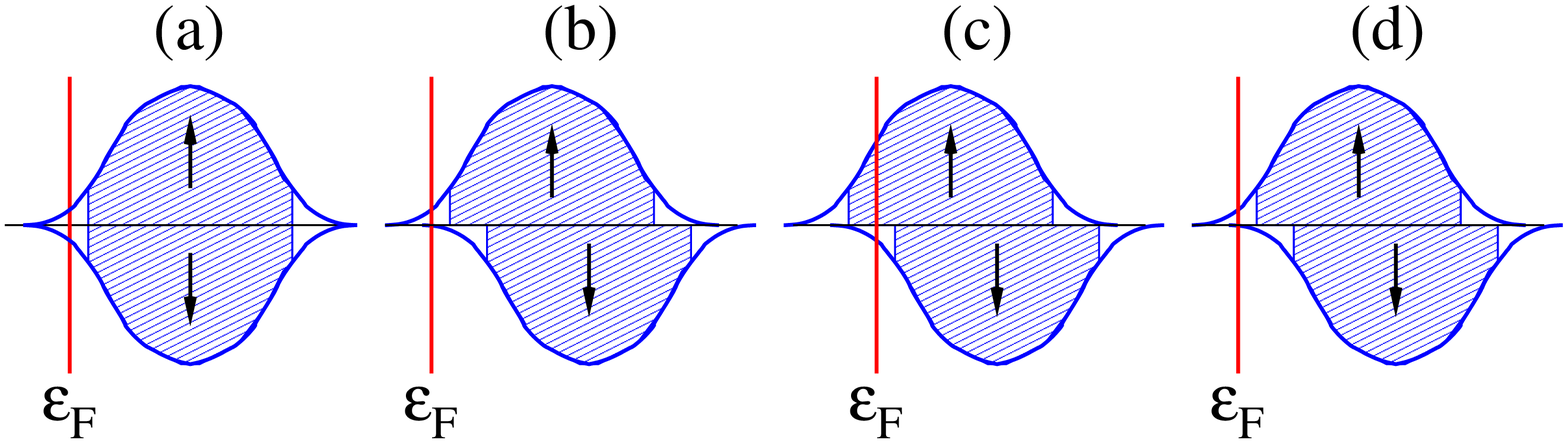}
\caption{Schematic density of states (DOS) versus energy for the impurity
band, for up- (top) and down-spins (bottom). Under each DOS curve, the  
hashed and empty regions correspond, respectively, to delocalized and
localized states; these are separated by mobility edges. The exchange
splitting is proportional to the off-set between the $\uparrow$ and
$\downarrow$ bands, and the Fermi energy ($\varepsilon_{\rm F}$) increases
to the right, towards the top of the valence band(not shown).}
\label{dos}
\end{figure}

At this point several comments are in order. First, the model is indeed
very simple, for it does not incorporate aspects such as a Kohn-Luttinger
treatment of the valence states\cite{Dietl01,Brum}, effects of impurity
potentials, a site energy term arising from the Mn potential, a
correlation energy representing hole-hole repulsion, and so forth. In
addition, the model is treated within a mean-field approximation which
neglects fluctuations in spin, charge, and disorder degrees of freedom.
Nonetheless, one expects these limitations to be minimized, to some
extent, by the fact that experimental data for $T_c(x)$ are used as input.
However, it is exactly this simplicity that allows us to obtain {\em a
direct} relation between hole concentration and Mn fraction, which, in
turn, can be promptly used as a rough guide to experiments.
Clearly, the present results must be viewed as a first approximation to
$p(x)$, since one should still be able to obtain such relation
phenomenologically through improved models and approximations, though with
a considerable amount of extra computational effort.  Consider, for
instance, the case of Monte Carlo simulations of the Kohn-Luttinger
Hamiltonian for the semiconductor valence bands\cite{Schliemann01}: since
the dependence of $T_c$ on $J_{pd}$ and on $m^*$ is different from that of
the mean-field prediction [Eq.\ (\ref{tc})], the one-parameter fitting
strategy adopted here is not so straightfowardly applicable; these
parameters would have to be separately adjusted, demanding many additional
runs. It is therefore hardly surprising that most of the improved
theoretical approaches\cite{Dietl01,Schliemann01,Chatt01,Jungwirth01}
consider $p$ (instead of $x$) as an independent variable and, accordingly,
present plots of $T_c(p)$, {\em for fixed $x$}; the issue of the relation
between $p$ and $x$ is then set aside.

The present approach also leads to a qualitative
understanding for the dependence of $p$ with $x$ being
essentially
related to the occurrence of MIT's taking place in the hole subsystem.
Within our approximation, the Fermi energy tracks the behaviour of $p,$
since $\varepsilon_{\rm F} \propto p^{2/3}$, while the exchange splitting
$\Delta\propto x$. Figure \ref{dos}(a) shows the schematic impurity bands
for each spin channel, in the very-low doping regime in which the gas is
supposed to be unpolarized.  As $x$ increases, the gas can sustain
polarization and still be insulating, provided the Fermi energy lies below
the mobility edge, as shown in Fig.\ \ref{dos}(b).  Further increase in
$x$ causes $\varepsilon_{\rm F}$ to increase and to lie within the
delocalized states of the up-spin impurity band, as depicted in Fig.\
\ref{dos}(c): The system becomes metallic. Whether or not the Fermi energy
also lies within the delocalized region of down-spin impurity band is a
very interesting question, which cannot be answered by our simple model;
the solution of this particular issue should have bearings on the
efficiency of Ga$_{1-x}$Mn$_x$As-based devices as spin filters. Once
$\varepsilon_{\rm F}$ reaches a maximum within the metallic phase, its
initial decrease upon increasing $x$ is compensated by an increase in
$\Delta$, so that the Fermi level still lies within the delocalized
states. However, with continuing increase in $x$ the exchange splitting
can no longer make up for the decrease in $\varepsilon_{\rm F}$, and the
latter eventually crosses the mobility edge again, lying within localized
states [Fig.\ \ref{dos}(d)]: The system reenters an insulating phase.  
One may argue that a description in terms of {\em impurity levels} rather
than {\em impurity bands} may be more adequate in the range of Mn
concentrations considered here. Even so, the movement of the Fermi energy
described above is still applicable with slight modifications: the
metallic phase would then correspond to $\varepsilon_{\rm F}$ reaching the
top of the valence band. This latter picture is actually in qualitative
agreement with recent photoemission measurements\cite{photo} of the Fermi
level as a function of Mn concentration in Mn$_x$Ga$_{1-x}$As.

\begin{figure}[t]
\epsfxsize=3.5in
\epsfbox{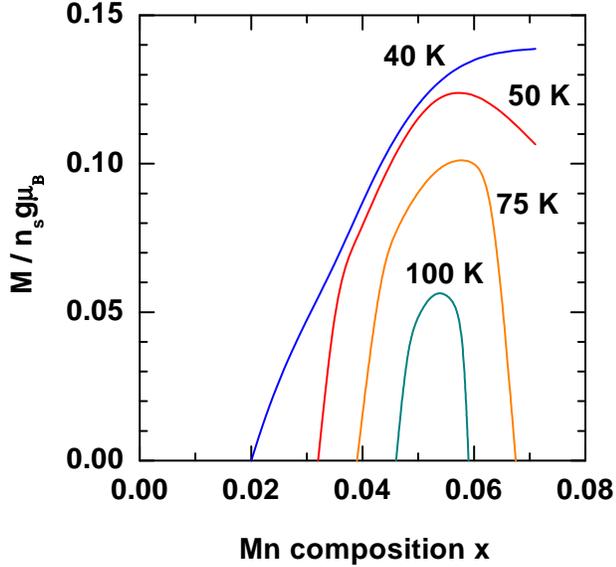}
\caption{Mn contribution to the magnetization [Eq.\
(\ref{M})] as a function
of the fraction of Mn sites, for four different temperatures, $T=40$ K, 50
K, 75 K, and 100 K.}
\label{mag}
\end{figure}

We now discuss the magnetization of the Mn ions, as obtained by solving
Eqs.\ (\ref{M}) and (\ref{mh}) for $M$ for a given Mn
composition and temperature. The mean-field theoretical results are shown in
Fig.\ \ref{mag} for various temperatures; for $T=75$ K and 100 K, we had to
resort to an interpolation of the experimental $T_c$ data from Ohno and
Matsukura\cite{Ohno01a}. Two effects are apparent from the calculated
results. Firstly, the magnetization maxima with respect to $x$ decrease with
increasing temperature, as it would be expected since one approaches the
critical temperature from below [see also Fig.\ 3(c) of Ref.\ 
\onlinecite{Ohno01a}]. Moreover, the widths of the composition-dependent
magnetization curves decrease quite drammatically with temperature. An
immediate consequence of these results is that DMS (Ga,Mn)As device
applications at temperatures $\lesssim T_c$ would require a definite
quality-control of the Mn doping composition.

In summary, we have established a theoretical scenario for the behaviour
of the hole concentration in Mn$_x$Ga$_{1-x}$As as a function of both $x$
and $T_c$, based on a simple mean-field approximation to the hole-mediated
ferromagnetic Hamiltonian. In our picture, the concentration of holes is
approximately constant in the low-doping insulating phase, then rises to a
maximum in the metallic phase, and drops again in the reentrant insulating
phase. Our approach also allows one to view the underlying mechanism of
the reentrant MIT's as an oscillation of the Fermi energy caused by a
delicate balance between band filling and exchange-splitting.  We have
also noted that the larger the temperature, the narrower is the range of
compositions leading to a non-zero Mn magnetization.  The present approach
should certainly be extended to include a more complete description of the
acceptor states, taking into account the spin degrees of freedom,
spin-orbit coupling, compressive/tensile strains, etc. Moreover, a proper
treatment of disorder -- e.g., by explicitly considering a random, instead
of continuous, distribution of Mn ions -- should lead to a more realistic
description of the MIT. In this respect, many-body effects due to
correlation among the holes should also influence $p(x)$, especially in
the insulating phase. Of course, an appropriate description of the
physical mechanisms related to As antisites and Mn$^{6{\rm As}}$ centres,
as far as the hole vs. Mn concentrations is concerned, is certainly a
formidable task, which nonetheless deserves future theoretical attention.

\bigskip 

The authors are grateful to J Schliemann for useful discussions, and to
the Brazilian Agencies CNPq, FAPESP, FAEP-UNICAMP, and FAPERJ for partial
financial support. RRdS and JdAC also gratefully acknowledged support
from the `Millenium Institute for Nanosciences/CNPq'.


\begin{references}

\bibitem{Ohno92}  H.\ Ohno, 
H.\ Munekata, T.\ Penney, S.\ von Moln\`ar, and L.\ L.\ Chang, 
\prl {\bf 68}, 2664 (1992).

\bibitem{Ohno96} H.\ Ohno, 
A.\ Shen, F.\ Matsukura, A.\ Oiwa, A.\ Endo, S.\ Katsumoto, and Y. Iye, 
\apl {\bf 69}, 363 (1996).

\bibitem{Ohno98a}  H.\ Ohno, Science {\bf 281}, 951 (1998).

\bibitem{Flederling99}  R.\ Flederling, 
 M.\ Kelm, G.\ Reuscher, W.\ Ossau,
G.\ Schmidt, A.\ Waag, and L.\ W.\ Molenkamp, 
Nature {\bf 402}, 787 (1999).

\bibitem{Ohno99a}  Y.\ Ohno, 
 D.\ K.\ Young, B.\ Beschoten, F.\ Matsukura,
H.\ Ohno, and D.\ D.\ Awschalom, 
Nature {\bf 402}, 790 (1999).

\bibitem{Ohno99c}  H.\ Ohno, J.\ Magn.\ Magn.\ Mater.\ {\bf 200}, 110 (1999).

\bibitem{Ohno00a}  T.\ Dietl, 
 H.\ Ohno, F.\ Matsukura, J.\ Cibert, and D.\ Ferrand, 
Science {\bf 287}, 1019 (2000).

\bibitem{DiV99} D.\ P.\ DiVincenzo, \jap {\bf 85}, 4785 (1999).

\bibitem{Hayashi97} T.\ Hayashi, 
 M.\ Tanaka, and T.\ Nishinaga, 
\jap {\bf 81}, 4865 (1997).

\bibitem{Ohno01a} H.\ Ohno and F.\ Matsukura, Solid State Commun.\ {\bf
117}, 179 (2001).

\bibitem{Ohno98b} F.\ Matsukura, 
 H.\ Ohno, A.\ Shen, Y.\ Sugawara, 
\prb {\bf 57}, R2037 (1998).

\bibitem{Ohno99b}  J.\ Szczytko, 
 W.\ Mac, A.\ Twardowski, F.\ Matsukura, and H.\ Ohno, 
\prb {\bf 59}, 12935 (1999).

\bibitem{Dietl01}  T.\ Dietl, 
H.\ Ohno, and F.\ Matsukura, 
\prb {\bf 63}, 195205 (2001).

\bibitem{Look91}  D.\ C.\ Look, J.\ Appl.\ Phys.\ {\bf 70}, 3148 (1991).

\bibitem{vanEsch97}  A.\ Van Esch, 
L.\ Van Bockstal, J.\ De Boeck, G.\
Verbanck, A.\ S.\ van Steenbergen, P.\ J.\ Wellmann, B.\ Grietens, R.\
Bogaerts, F.\ Herlach, and G.\ Borghs, 
\prb {\bf 56}, 13103 (1997).

\bibitem{Dietl97} T.\ Dietl, 
 A.\ Haury, and Y.\ Merle d'Aubign\'e, 
\prb {\bf 55}, R3347 (1997).

\bibitem{Mac99a}  T.\ Jungwirth, 
 W.\ A.\ Atkinson, B.\ H.\ Lee, and A.\ H.\ MacDonald, 
\prb {\bf 59}, 9818 (1999).

\bibitem{Mac00a} B.\ Lee, 
 T.\ Jungwirth, and A.\ H.\ MacDonald, 
\prb {\bf 61}, 15606 (2000).

\bibitem{Quinn00} S.\ P.\ Hong, 
K.\ S.\ Yi, and J.\ J.\ Quinn, 
\prb {\bf 61}, 13745 (2000).

\bibitem{Mac00b}  J.\ K\"onig, 
H.-H.\ Lin, and A.\ H.\ MacDonald,
cond-mat/0010471.

\bibitem{Litvinov01} V.\ I.\ Litvinov and V.\ K.\ Dugaev, \prl {\bf 86},
5593 (2001).

\bibitem{Ohno99d}  H.\ Ohno, 
 F.\ Matsukura, T.\ Omiya, and N.\ Akiba, 
J.\ Appl.\ Phys.\ {\bf 85}, 4277 (1999).

\bibitem{foot1}  As pointed out in Ref.\ \onlinecite{Ohno99c}, the currently
accepted value $p=3.5\times 10^{20}$ cm$^{-3}$ is a factor 2.3 larger than
the early estimate of Ref.\ \onlinecite{Ohno98b}.

\bibitem{foot2}  For instance, Fig.\ 2 of Ref.\ \onlinecite{Ohno98b}
indicates that for $x=0.022$, the experimentally determined $p$ has an error
bar which covers over one decade, while for $x=0.071$ the error bar runs
over two decades.

\bibitem{photo} H.\ Asklund, L.\ Ilver, J.\ Kanski, and J.\ Sadowski,
cond-mat/0112287.

\bibitem{Brum}  M.\ Abolfath, 
T.\ Jungwirth, J.\ Brum, and A.\ H.\ MacDonald, 
\prb {\bf 63}, 054418 (2001).

\bibitem{Schliemann01} J Schliemann, J K\"onig, and A H MacDonald, \prb
{\bf 64}, 165201 (2001).

\bibitem{Chatt01} A.\ Chattopadhyay, S.\ Das Sarma, and A.\ J.\ Millis,
\prl {\bf 87}, 227202 (2001).

\bibitem{Jungwirth01} T Jungwirth, B Lee, and A H MacDonald, Physica
E {\bf 10}, 153 (2001). 




\end{references}
\end{document}